# Do feelings matter? On the correlation of affects and the self-assessed productivity in software engineering

**Daniel Graziotin, Xiaofeng Wang, Pekka Abrahamsson**
**Free University of Bozen-Bolzano, Bolzano, Italy**
**{daniel.graziotin, xiaofeng.wang, pekka.abrahamsson}@unibz.it**

ABSTRACT

**Background**: software engineering research (SE) lacks theory and methodologies for addressing human aspects in software development. Development tasks are undertaken through cognitive processing activities. Affects (emotions, moods, feelings) have a linkage to cognitive processing activities and the productivity of individuals. SE research needs to incorporate affect measurements to valorize human factors and to enhance management styles.

**Objective**: analyze the affects dimensions of valence, arousal, and dominance of software developers and their real-time correlation with their self-assessed productivity (sPR).

**Method**: repeated measurements design with 8 participants (4 students, 4 professionals), conveniently sampled and studied individually over 90 minutes of programming. The analysis was performed by fitting a linear mixed-effects (LME) model.

**Results**: valence and dominance are positively correlated with the sPR. The model was able to express about 38% of deviance from the sPR. Many lessons were learned when employing psychological measurements in SE and for fitting LME.

**Conclusion**: this article demonstrates the value of applying psychological tests in SE and echoes a call to valorize the human, individualized aspects of software developers. It reports a body of knowledge about affects, their classification, their measurement, and the best practices to perform psychological measurements in SE with LME models.

## 1. INTRODUCTION

For more than thirty years, people have claimed that a way to improve software developers' productivity and software quality is to focus on people and make them satisfied and happy [1]. Several Silicon Valley companies and software startups follow this claim by providing so called "perks" to make their employees happy [2,3,4] and, allegedly, more productive [5].

Examples of these "perks" include assigning them private offices, creating a working environment to support creativity, and providing incentives [1].

Human aspects play an important role in the execution of software processes and the resulting products [6,7,8]. However, human aspects have been neglected for a long time in managing developers and when conducting research on software development. This perception of the importance of the human aspects in software development - e.g., "Individuals and interactions over processes and tools" - led to the publication of the Agile manifesto [9]. Advocates of agile software development added that "If the people on the project are good enough, they can use almost any process and accomplish their assignment. If they are not good enough, no process will repair their inadequacy – 'people trump process' is one way to say this." [10:1].

There is an increasing awareness in software engineering research that links human-related factors to software development productivity and performance [11][1]. But, how can research in empirical software engineering consider human aspects? One proposal is by using psychometrics in empirical software engineering studies [12].[2] Although it has been established that software development is carried out through intellectual undertakings, thus cognitive processing activities [6,7,13,14], research in software engineering has forgotten that affects - emotions, moods, and feelings - have an impact on the cognitive activities of individuals [14].

As individuals, we act based on emotions as we discover the world through a certain mood. Ciborra put it well by stating that affects enable the "mattering" of things; they are the medium in which acting towards the world takes place [15:159–165]. The role of affective states (or affects) in the workplace has received significant attention in management and psychology research [16,17,18]. In psychology research, some evidence has been found to prove that happier employees are more productive [17,18,19]. Other studies instead suggested that affects relate to job performance differently with respect to the various aspects or types of jobs [20]. However, still little is known about the productivity of individual programmers [21] and the link between affects and the performance of developers [14,22]. It is necessary to understand how affects play a role in software development as real-time correlations of performance and affects are often overlooked [19,23].

Deng [24] recently argued that the recognition moods of developers is essential to professional success in software development. There are also several calls for research on the role of emotions in software engineering [8,14,22,24]. However, the actual research in the field is very limited. For example, research in software process improvement began in the last decade to consider the commitment of people a necessary factor to determine the success of well-planned process improvement programs. Some claim that affects in the form of affective commitment are among the necessary factors to succeed in well-planned software process improvement initiatives [25]. Apart from the little input in affect-related factors, there is a limited knowledge on how affects play a role in the performance of software developers. This manuscript is a contribution towards a better understanding of the real-time role of affects during software development.

---

[1] The stance that performance and productivity are two interchangeable terms is assumed in this study, in line with [20,84,85] from both software engineering and psychology research.

[2] The software engineering literature has sometimes misused the term *psychometrics* to describe metrics derived from psychology. However, *psychometrics* has a specific meaning within psychological research and involves establishing the reliability and validity of a psychological measurement. In this article, we use the more appropriate term of *psychological measurement* to refer to metrics derived from psychology.

The *research question* that this study aims to answer is "How are the affects related to a software development task correlated with the immediate, self-assessed productivity of developers?". To this end, this study examines the variations of affects and the self-assessed productivity of software developers while they are programming.

The research *objective* expressed with the goal template of the Goal/Question/Metric method [26], as explicated by Wohlin et al. [27], is to analyze the self-assessed productivity for the *purpose* of evaluation *with respect to* the affects' dimensions (valence, arousal, dominance) from the *point of view* of the researcher in the *context* of Computer Science students and professional developers programming on their software projects. The context of the experiment is natural settings, i.e., the working environment.

The *main results* of the study are threefold. 1) The affect dimensions of valence and dominance of software developers are positively correlated with their self-assessed productivity. 2) The investigation produced evidence on the value of psychological testing in empirical software engineering studies and the appropriate analysis methods, i.e., linear mixed effects models, to be employed within participant repeated measures. 3) The article reports a substantial body of knowledge in psychology and management research on the theories behind emotions, their classification, their measurement, and the best practices to perform psychological measurements in the context of empirical software engineering.

This article is an extension of a conference article presented at PROFES 2013 [28]. This article adheres to the proposed guidelines when reporting experiments in software engineering [29] and is structured as follows. Section 2 provides the background theory on human aspects in productivity research, the affects and their measurement, and the related studies on the correlation between affects and the performance of individuals. Section 3 describes all the possible details of the research methodology in order to be easily evaluated and replicated. Section 4 merges the proposed Execution and Analysis sections; it reports how the experimental plan was enacted and how the data was analyzed. The section ends with the hypotheses testing. Section 5 discusses the obtained results, their implications, the limitations of the study, and the lessons learned. Section 6 concludes the article with a summary, the impact of this study, and suggestions for future work.

# 2. RELATED WORK

## *2.1 Background*

### *2.1.1 Human aspects in productivity studies*

This subsection reviews the major studies on productivity factors in software engineering. They are reported in chronological order so that the refinement of the awareness of human aspects in the performance of software developers is highlighted.

More than 30 years ago, programmer's productivity was believed to be influenced by different characteristics. Chrysler [30] considered developers' productivity to be influenced by characteristics at the technical level, the knowledge level, and the developer level. However, only skills and experience, measured as the number of months, were taken into consideration as human factors.

Boehm and Papaccio [1] identified factors influencing productivity by controlling the costs of producing software. The study suggested strategies to improve productivity such as writing less code and "getting the best from people". This study suggested ways for improving productivity such as assigning people private offices, creating a working environment to support creativity, and providing incentives to enhance the motivation of people and their commitment.

Scacchi [21] observed which factors influenced software productivity and how productivity can be improved. His review focused on the creation of a framework to predict the productivity of large-scale software systems. The study criticized the previous research, because they failed to describe the variation in productivity among individual programmers. An important impact was attributed to human-related factors. Scacchi argued that organizational and social conditions can even dominate the productivity attributable to in-place software development technologies. The study called for improvement and alternative directions in software productivity measurements.

A recent review of productivity factors by Sampaio et al. [11] identified three main areas in the body of knowledge: product, project, and people. The identified people-related factors consisted of the motivation of the team, and the individuals' skills, but also relationships, and the quality of management.

### *2.1.2 Affects, emotions, moods, and feelings*

The term *affective states* (affects) has been defined as "any type of emotional state [...] often used in situations where emotions dominate the person's awareness" [31]. However, as is shown in this section, affects have been employed as a general term for emotions, moods, and feelings.

*Emotions* have been defined as the states of mind that are raised by external stimuli and are directed toward the stimulus in the environment by which they are raised [32]. However, more than 90 definitions have been produced for this term [33], and no consensus within the literature has been reached. This term has been taken for granted and is often defined with references to a list, e.g. anger, fear, joy, surprise [34]. *Moods*, on the other hand, have been defined as emotional states in which the individual feels good or bad, and either likes or dislikes what is happening around him or her [35]. *Feelings* have been defined as the conscious subjective experience of emotions [31]. The term *happiness* has been defined as the emotional evaluation of life [36,37] measured as the sum of the frequency of emotions in a timespan [37,38].

The difference between these terms is subtle for researchers outside of the psychology fields, making it common for these researchers to interchange the terms in everyday life. Indeed, many authors over time have considered mood, emotion, and feeling as interchangeable terms [39,40,41,42], although it has been acknowledged that numerous attempts exist to differentiate them [42,43]. Some scholars suggested that the difference between moods and emotions lies in an absence of a causal factor in the phenomenal experience of the mood [43]. Many researchers state that emotions and moods are affective states [14,17,43,44]. Some have claimed that the distinction is not truly necessary for the sake of studying cognitive responses that are not strictly connected to the origin of the mood or emotion [43]. For the purposes of this investigation, the authors have adopted the latter stance and have employed the noun *affective states* as an umbrella term for emotions, moods, and feelings.

There are two main theories to categorize affects. One theory, called the discrete approach, seeks a set of basic affects that can be distinguished uniquely [32] and possess high cross-cultural agreement when evaluated by people in literate and preliterate cultures [45,46]. Ekman [45] proposed a sets of basic emotions, which include anger, happiness, surprise, disgust, sadness, and fear. However, the list received critiques and it was extended with eleven other emotions [47]. These newly added emotions include amusement, embarrassment, relief, and shame. Plutchik [32] proposed an alternative viewpoint, called the Wheel of Emotions. Eight primary, bipolar emotions are coupled: joy versus sadness, anger versus fear, trust versus disgust, and surprise versus anticipation. These eight basic emotions can vary in intensity and can be combined with one other to form secondary emotions. For example, joy is the midpoint between serenity and ecstasy, whereas sadness is the midpoint between pensiveness and grief [32]. A list of basic affects is difficult to keep minute in size with the discrete approach as some studies proposed more than one hundred basic emotions [48].

The other theory of emotion groups affects in major dimensions, which allow a clear distinction among them [49,50]. In this theory, affective states are characterized by their valence, arousal, and dominance. *Valence* (or pleasure) can be described as the attractiveness (or adverseness) of an event, object, or situation [51,52]. It refers to the "direction of a behavioral activation associated toward (appetitive motivation) or away (aversive motivation) from a stimulus" [50:990]. *Arousal* represents the intensity of emotional activation [50]. It is the sensation of being mentally awake and reactive to stimuli – i.e. vigor and energy or fatigue and tiredness [53]. *Dominance* (or control, over-learning) represents a change in the sensation of control of a situation [54]. It is the sensation by which an individual's skills are perceived to be higher than the challenge level for the task [55]. The dimensional approach distinguishes itself from the discrete approach in its fewer elements to evaluate. Thus, this theory is deemed useful in tasks where affects must be evaluated quickly and preferably often. The dimensional approach is commonly adopted to assess affects triggered by an immediate stimulus [54,56] making it common in human-machine interaction and computational intelligence studies (e.g., [57,58]). Because of these reasons, the dimensional approach was adopted in this study.

#### *2.1.3 Measuring affects*

The measurement of emotions has often been achieved using surveys. One of the most used questionnaires for the dimensional approach is the Self-Assessment Manikin (SAM) [54,59]. SAM is a non-verbal assessment method based on pictures. SAM measures valence, arousal, and dominance associated with a person's affective reaction to a stimulus [54]. A numeric value is assigned to each rating scale for each dimension. A screenshot of this study measurement instrument is provided in the online appendix of this study [60] and it provides a representation of the SAM items. SAM is not uncommon in computer science research where the affects towards a stimulus has to be studied (e.g., [57]). SAM was chosen as the measurement instrument for this study because of its peculiarities, e.g., the pictorial items. The shortness of the questionnaire itself, along with the fact that it is commonly employed in other fields including computer science, and the about 2000 studies either referencing it or employing it were also taken into consideration when this method was chosen for this study.

These scales, and similar other psychological measures, present issues when employed within- (and between-) subjects analyses of repeated measurements which is the case of this study. First, there is not a stable and shared metric for assessing the affects across persons. For example, a score of one in valence for a person may be equal to a score of three for another person. However, a participant scoring two for valence at time $t$ and five at time $t+x$ unquestionably indicates that the participant's valence increased. Therefore, as stated by Hektner et al. [61], "it is sensible to assume that there is a reasonable stable metric within persons" (p. 10). In order to have comparable measurements, the raw scores of each participant are typically transformed into *z-scores* (also known as standard scores). The *z-score* transformation is such that a participant's mean score for a variable is zero, and scores for the same variable that lie one standard deviation above or below the mean have the value equivalent to their deviation. Therefore, one observation is expressed by how many standard deviations it is above or below the mean of the whole set of an individual's observations. In this way, the measurements between participants become dimensionless and comparable with each other [61,62], as they indicate how much the values are spread.

Second, the repeated measurements employed in contexts, like the one of this study, present dependencies of data at the participants' level and the time level grouped by the participant. The *ANOVA* family provides *rANOVA* as a variant for repeated measurements. However, *rANOVA* and general *ANOVA* procedures are discouraged [63] in favor of mixed-effects models which are robust and specifically designed for repeated, within-participant longitudinal data [63,64,65].

### *2.2 Related studies*

This section reviews the studies directly related to this work. In the first part, the article presents the prominent studies in psychology and management where the affects of workers are studied with respect to their performance and productivity. In the second part, the paper presents the few papers in the area of software engineering that study the affects of software engineers.

First, it is not uncommon in psychology and management research to let participants self-assess their productivity [18,19,23], as self-assessed performance is a consistent (yet not preferred) method to objective measurements of performance [66,67]. There is also evidence that bias is not introduced by mood and emotions in self-reports of performance, especially when individuals alone are being observed [23,66].

Fisher and Noble [19] employed the Experience Sampling Method [62] to study correlations between real-time performance and affects while working. The study recruited different workers (e.g., childcare worker, hairdresser, office worker); however, none of these workers were reported to be a software developer. The measurement instrument was a questionnaire with 5-point Likert items. The article analyzed self-assessed skills, task difficulty, affects triggered by the undertaken task, and task performance. The results of the study revealed that there was a strong positive correlation between positive affects and task performance while there was a strong negative correlation between negative affects and task performance. This article encourages further research about within-subjects real-time performance and emotions.

Along the same line, Miner and Glomb [66] performed a similar study with 67 individuals working in a call-center. The individuals were assessed 5 times per day. Within-subjects, the periods of positive mood were found to be associated with the periods of improved task performance, in this case in terms of shorter support call-time.

Oswald, Proto, and Sgroi [17] argued that research in management has lacked studies on the relationship between happiness of workers and their productivity. They conducted a controlled experiment where 182 participants were divided into two groups. The first group received treatment of positive affects induction – i.e., a comedy clip, while the second group did not receive any treatment. The participants performed two mathematical tasks and their performance in the tasks represented their productivity. The results showed that a rise in positive affects leads to higher productivity. The effect was found to be equally significant in both male and female subsamples.

Shaw [22] observed that the role of emotions in the workplace has been the subject of management research. However, little or no attention has been given to the emotions of Information Technology (IT) professionals. He conducted an exploration of the emotions of IT professionals. The study focused on how emotions can help to explain IT-related job outcomes. The paper employed the Affective Events Theory [43] as a framework for studying the fluctuation of the affects of twelve senior-level undergraduate students working on a semester-long implementation project for a university course. The participants were asked to rate their affective states during or right after their episodes of work on their projects. Shaw considered each student a single case study, as statistical analysis was not considered suitable. The study showed that the affective states of a software developer may dramatically change during a period of 48 hours. However, the research was a work in progress, and no continuation of the project is currently known. The paper has called for research on the affective states of software developers.

Lesiuk [68] performed a quasi-experimental field study with an interrupted time series with removed treatment. She recruited 56 software engineers, who were working in four software companies, to understand the effects of music listening on software design performance. Data was collected over a five-week period, where the performance was self-assessed twice per day by the participants, together with their affective states. For the first week of study (the baseline) the participants were only observed in natural settings. During the second and third week, the participants were allowed to listen to their favorite music whenever they preferred. During the fourth week, the software engineers were not allowed to listen to any music, all day long. During the fifth week, the participants were allowed again to listen to the music. The results indicated that positive affects are positively correlated with music listening (or better, with the allowance of music listening). Then, positive affects of the participants and self-assessed performance were lowest with no music, while time-on-task was longest when music was removed. However, the correlation was not statistically significant. Narrative responses revealed the value of music listening for positive mood change and enhanced perception on design while working.

Khan et al. [14]provided links from psychology and cognitive science studies to software development studies. The authors constructed a theoretical two-dimensional mapping framework in two steps. In the first step, programming tasks were linked to cognitive tasks. For example, the process of constructing a program – e.g. modeling and implementation – was mapped to the cognitive tasks of memory, reasoning, and induction. In the second step, the same cognitive tasks were linked to affects. Two empirical studies on affects and software development were conducted, which related a developer's debugging performance to induced affects. In the first study, affects were induced to software developers, who were then asked to complete a quiz on software debugging. The second study was a controlled experiment.

The participants were asked to write a trace of the execution of algorithms implemented in Java. The results suggest that (1) induced high valence condition alone does not coincide with high debugging performance, (2) induced high arousal condition alone coincides with high debugging performance, and (3) induced high arousal and valence conditions together are associated with high debugging performance. This study recommended more research on the topic.

Colomo-Palacios et al. [8] considered requirements engineering as a set of knowledge-intensive tasks. The study aimed to integrate the stakeholder's emotions into the process of requirements engineering. The authors conducted two empirical studies on two projects. The first project consisted in the maintenance of a legacy system, while the second project was the development of a touristic information system. In total, 65 user requirements were produced between the two projects which lasted between six and seven months respectively. Each requirement faced different revisions, up to 97 for the first project and up to 115 in the second project. Each participant rated the affective state associated to each requirement version. Affective states in this study were represented with the dimensional state using the components of valence and arousal. The results showed that high arousal and low pleasure levels are predictors of high versioning requirements. Additionally, valence increased throughout versions (thus, over time), while the arousal decreased. The authors questioned what could be the role of time in the emotional rating of the participants, and called for more research on the role of affects in software engineering.

Wrobel [69] conducted a survey with 49 developers. The questionnaire assessed the participants' job-related affects during programming, and which emotions were perceived to be those influencing their productivity. The results showed that developers feel a broad range of affects while programming—all the affects of the measurement instrument's spectrum. positive emotions dominate in their work. The five most frequently occurring emotional states were happy, content, enthusiastic, optimistic, and frustrated. That is, the four most experienced emotions were positive. Positive affects were perceived to be those enhancing their productivity. It is interesting to note that 13% of the developers indicated a positive or very positive impact on productivity when they were angry. However, the result was not statistically significant overall. Finally, frustration was perceived as the negative affect more often felt, as well as the one perceived as deteriorating productivity.

Graziotin et al. [70] echoed the call for research on alternative factors influencing the performance of software developers. They conducted a study with 42 computer science students to investigate the relationship between the affects and creative and analytical performance of software developers. The participants performed two tasks coming from psychology research. The first task was related to creative performance, the other was related to analytic performance and resembled algorithm design and execution. The participants' pre-existing affects were measured before each task. The analysis of the data showed empirical support for the claim that happy developers are indeed better problem solvers in terms of their analytical abilities. The study raised the need for studying the human factors of software engineering by employing a multidisciplinary viewpoint.

## 3. EXPERIMENT PLANNING

### *3.1 Participants*

The participants for this experiment are software developers. Professionals and students can both be taken into consideration for being participants. The only requirement is that the participants work on any software project, but can perform a task individually. There are no restrictions in gender, age, or nationality of the participants. Participation is voluntary and not rewarded. The rationale is that participants work on their project in their natural settings. Therefore, they only need to accept the presence of the researcher for a limited time. Confidentiality has to be assured to participants upfront when they are recruited. They are asked to participate in a study in which they are singularly observed for a limited amount of time and asked to fill a very short questionnaire at regular intervals of ten minutes during work. They are assured on the anonymity of the gathered data as well. No personal information is retained.

### *3.2 Experimental material*

On the participant side, there is no required material. Participants develop software in natural settings using their own equipments – i.e., a computer. The researcher, however, needs a suitable tablet device that implements the questionnaire to measure the affects of the participants and their self-assessed productivity. This study employed an ad-hoc Web-based survey, available upon request. Since the questionnaire is pictorial (See [60]), the effort required for a measurement session is reduced to four touches to the screen.

This experiment involves pre- and post-task interviews and the annotation of events during the task execution. Therefore, suitable devices for the recording of the observations are also required, i.e., a notepad and an audio recorder.

### *3.3 Procedure*

Figure 1 summarizes the entire procedure for this study, as a timeline. The procedure is composed of three parts: a pre-task interview, a software development task, and a post-task interview.

As indicated in the left part of Figure 1, the researcher and the participants first engage in a pre-task interview. The starting questions for the pre-task interview can be found in the online appendix of this study [60]. The participants are interviewed either in natural settings – e. g., their offices – or in any location that makes them comfortable, like a bar. During the pre-task interview, basic demographic data, information about the project, tasks, and the developers' skills are obtained. Descriptive data is collected on the role of the participant (either "professional" or "student"), as well as the perceived experience with the programming language, and the perceived experience with the domain of knowledge – including the task type - (low, medium, and high).

Right after the pre-task interview, as indicated by the central part of Figure 1, the participants and the researcher sit together in the working environment. Before the start of the task, the participants are informed about the mechanisms involved in the task session. The instructions given to each participant are available in the on-line appendix of this article [60]. The instructions for the Self-Assessment Manikin (SAM) questionnaire were written following the technical manual by Lang et al. [59]. For a period of 90 minutes, the participants work on software development tasks of real software projects. The researcher observes the behavior of the individuals while they are developing software. After each 10 minute session, the participants complete a short questionnaire on a tablet device. In this questionnaire, valence, arousal, and dominance are measured 9 times per participant. The same process holds for self-assessment of the worker's productivity. The researcher is present during the entire development period to observe the behavior of the participant without interfering.

After the completion of the working period, the researcher conducts a post-task interview represented by the third part of Figure 1. The questions are available in the online appendix of this study [60] and are related to the self-assessment of the productivity of the participants, the factors influencing the performance, and if and how the measurement system interrupts or annoys the participants. The interviews are structured and mostly close-ended questions are used. The purpose is to triangulate the findings obtained in the main experiment.

The reader is reminded that this experiment studies the participants one at a time. Each step of the experiment procedure is executed on an individual basis. Additionally, all steps of the experiment are automatized through electronic systems.

### *3.4 Goals, hypotheses, parameters, and variables*

The body of knowledge presented in section 2 suggests that affects are positively correlated to the productivity of individuals. Therefore, the research hypotheses of this study are on positive correlations between real-time affects and the self-assessed productivity of software developers. Figure 2 builds upon Figure 1, to explain the formulation of the hypotheses and the involved factors in the context of the experimental design[3].

H1. The real-time valence affective state of software developers is positively correlated to their self-assessed productivity.
H2. The real-time arousal affective state of software developers is positively correlated to their self-assessed productivity.
H3. The real-time dominance affective state of software developers is positively correlated to their self-assessed productivity.

[3] The authors are thankful to an anonymous reviewer for suggesting the graphical representations in Figure 1 and Figure 2.

The affects dimensions - valence, arousal, dominance - describe differences in affective meanings among stimuli and are measured with the SAM pictorial questionnaire. The values of the affective state constructs range from one to five. A value of three means “perfect balance” or “average” between the most negative (1) and the most positive value (5). For example, a value of one for the valence variable means “complete absence of attractiveness”. The task productivity is self-assessed by the participant, using a five-point Likert item. The item is the sentence “My productivity is ...” The participant ends the sentence, choosing the proper ending from the set {very low, below average, average, above average, very high}.

However, as reported in section 2.1.3, a participant’s data is converted to the individual’s *z-score* for the set of construct measurements, using the formula in (1):

$$z_{score}\left(x_{pc}\right) = \frac{x_{pc} - \bar{x}_{pc}}{s_{pc}} \qquad (1)$$

where $x_{pc}$ represents a participant’s measured construct, $\bar{x}_{pc}$ is the average value of all the construct measurements of the participant, and $s_{pc}$ is the standard deviation for the participant’s construct measurements. The reader is advised to read section 2.1.3 for more information about this transformation.

As the variables are transformed to z-scores, their values will follow a normal distribution in their range. The three-sigma rule states that 99.73% of the values lie within three standard deviations of the mean in a normal distribution [71]. Therefore, the range of the variables, while theoretically infinite, is practically the interval [-3, +3].

### *3.5 Analysis procedure*

Figure 1 and Figure 2 show that this experiment design is based on repeated measurements of multiple variables, with particular data dependencies at the participant’s level and at the time level. Linear mixed-effects models are the most valuable tool to be employed in such cases. A linear mixed-effects model is a linear model that contains both fixed effects and random effects. The definition of a linear mixed-effects model by Robinson [72] is given in (2):

$$y = X\beta + Zu + \varepsilon \qquad (2)$$

where $y$ is a vector of observable random variables, $\beta$ is a vector of unknown parameters with fixed values (i.e., fixed effects), $u$ is a vector of random variables (i.e., random effects) with mean $E(u) = 0$ and variance-covariance matrix $var(u) = G$, $X$ and $Z$ are known matrices of regressors relating the observations $y$ to β and $u$ , and $\varepsilon$ is a vector of independent and identically distributed random error terms with mean $E(\varepsilon) = 0$ and variance $var(\varepsilon) = 0$.

The estimation of the significance of the effects for mixed models is an open debate [73,74]. One proposed way is to employ likelihood ratio tests (*ANOVA*) as a way to attain *p-values* [75]. With this approach, a model is constructed by adding one factor (or interaction) at a time and performing a likelihood ratio test between the null model and the one-factor model. By adding one factor or interaction at a time, it would be possible to construct the best fitting possible model. However, this technique is time-consuming and prone to human errors.

Another proposed approach is to construct the full model instead. A way to express the significance of the parameters is to provide upper and lower bound *p-values*. Upper-bound *p-values* are computed by using as denominator degrees of freedom the number of data points minus the number of fixed effects and the intercept. Lower bound (or conservative) *p-values* are computed by using as denominator degrees of freedom the number of data points minus the within-participant slopes and intercepts multiplied per the number of participants. The reader is advised to read the technical manual [76] for additional details. This is the approach that was followed in this study.

The model has been implemented using the open-source statistical software *R* [77] and the *lme4.lmer* function for linear mixed-effects models [78]. For the model construction, valence, arousal, dominance, and their interaction with time are modeled as fixed effects. The random effects are two: a scalar random effect for the participant-grouping factor (i.e., each participant) and a random slope for the measurement time, indexed by each participant. In this way, the dependency of the measurements within participants is taken into account at the participant's level and at the time level. The full model is given in (3) as a *lme4.lmer* formula.

$$\text{productivity} \sim (\text{valence} + \text{arousal} + \text{dominance}) * \text{time} + (1|\text{participant}) + (0 + \text{time}|\text{participant}) \quad (3)$$

where *productivity* is the dependent variable; *valence*, *arousal*, *dominance*, and *time* are fixed effects; (1 | *participant*) is the scalar random effect for each participant, and (0 + *time* | *participant*) is read as "no intercept and time by participant"[4]. It is a random slope for the measurement time, grouped by each participant.

# 4. EXECUTION AND ANALYSIS

## *4.1 Preparation and deviations*

In this instance of the study, the participants were students of computer science at the Free University of Bozen-Bolzano and workers at local IT companies. The participants have been obtained using convenience sampling. However, as it is described in section 4.3, the sample provides a fair balance in terms of knowledge and roles. As the participants work on their own software projects, no particular training was needed. However, participants were trained on the measurement instrument using the supplied reference sheet [60]. No deviations occurred. As no dropouts happened and no outliers could be identified, the only required data transformation was the formula in (1). The *z-score* transformation was implemented in a *R* script with the command *scale()*.

[4] The authors are thankful to an anonymous reviewer, who correctly suggested this slight change to the original model ending with (time|participant).

### 4.2 Descriptive statistics

This study recruited eight participants for 72 measurements. The mean of the participants' age was 23.75 (standard deviation=3.29). Seven of them were male. Four participants were first year B.Sc. computer science students and four of them were professional software developers. The computer science students worked on course-related projects. The four professional software developers developed their work-related projects.

#### 4.2.1 Pre-task interviews

The characteristics of the participants, gathered from the pre-task interviews, are summarized in Table 1. A first observation is that the roles did not always correspond to the experience. The professional participant P2 reported a LOW experience with the programming language while the student participant P8 reported a HIGH experience in both the programming language, the domain of knowledge, and task type. Table 1 also contains the characteristics of the projects and the implemented task. There were a high variety of project types and tasks. Five participants programmed using C++ while two of them with Java and the remaining one with Python. The participants' projects were non-trivial, regardless of their role. For example, participant P1 (a professional developer) was maintaining a complex software system to collect and analyze data from different sensors installed on hydrological defenses (e.g., dams). Participant P5 (a student) was implementing pictorial exports of music scores in an open-source software for composing music.

#### 4.2.2 Repeated measurements

Figure 3, Figure 4, and Figure 5 provide the charts representing the changes of the self-assessed productivity (dashed line) over time, with respect to the valence, the arousal, and the dominance dimensions (solid line) respectively.

As seen in Figure 3[5], there were cases in which the valence score showed very close linkage to productivity (participants P2, P7, and P8). For the other participants, there were many matched intervals, e.g. P5 at interval 7, and P4 at intervals 4-7. Participant P1 was the only one for which the valence did not provide strong predictions. In few cases, the valence *z-score* was more than a standard deviation apart from the productivity *z-score*.

The arousal dimension in Figure 4[5] looked less related to the productivity than the valence dimension. The behavior of the arousal line often deviated from the trend of the productivity line (e.g., all the points of participants P5 and P6). Nevertheless, there were intervals in which the arousal values were closely related to productivity, e.g., with participants P4 and P7.

The dominance dimension in Figure 5[5] looked more correlated to the productivity than the arousal dimension. Participants P1, P5, and P7 provided close trends. For the other cases, there were intervals in which the correlation looked closer and stronger. However, it became weaker for the remaining intervals (e.g., with P4). The only exception was with participant P6 where a clear correlation between dominance and productivity could not be spotted.

[5] Figure 3, Figure 4, and Figure 5 are reprinted from [28] and reproduced here with kind permission from Springer Science and Business Media.

For all the participants, the *z-score* of the variables showed variations of about two units over time. That is, even for a working time of 90 minutes, there were strong variations of both the affects and the self-assessed productivity.

*4.2.3 Post-task interviews*

The post-task interviews were analyzed in light of the obtained results presented in the previous section. For the question regarding their satisfaction with their task performance, three participants (P4, P5, and P6) replied with a negative answer while the others answered with a nearly positive or completely positive answer.

All the participants had a clear idea about their productivity scores. None of them raised a question about how to self-rate their productivity. However, in the post-task interview, none of them was able to explain how they defined their own productivity. Six of them suggested that the self-assessment was related to the achievement of the expectation they set for the end of the 90 minutes of work. Their approach was to express the sequent productivity value with respect to the previous one – as in "worse", "equal", or "better" than before.

When questioned about difficulties and about what influenced their productivity, the participants found difficulties in answering in the beginning. Seven participants reported difficulties of technical nature. For example, P2 was slowed down because of "difficulties in finding the right functions of Python for scraping data from non-standard representations in the files". P4 was slowed down because of "an obscure bug in the Secure Socket Layer library" which prevented a secure communication channel to be opened. Only P1 complained about non-technical factors. P1 was interrupted twice by two phone calls from a senior colleague, between interval 3 and interval 5. The phone calls were related to P1's task, as he was asked to perform "urgent maintenance on related stuff" that he was working on. This was reflected by P1's self-assessed productivity. It is noted here that no participants mentioned affective-related factors when answering this question. When directly inquired about the influence of their affects on the performance of their development task, six participants responded negatively: they were convinced that affects did not have an impact on their task performance.

Of the eight participants, none of them reported to be annoyed or disturbed at all by the way the measurements were obtained. This was probably achieved because a series of pilot studies with other participants had been conducted in order to reduce the invasiveness of the measurement sessions. However, two participants suggested that a wider measurement interval would have been more welcome. For the question "Would it annoy you if this system was employed during the whole working day?," all the participants agreed that a measurement interval of 10 minutes would most likely reduce their productivity.

### *4.3 Hypotheses testing*

When fit with the gathered data, the proposed full model in (3) significantly performed better from its corresponding null model (4) in terms of likelihood ratio tests (*anova* in *R*; $\chi^2(7)=30.88$, $p<0.01$).

$$\text{productivity} \sim 1 + (1|\text{username}) + (0 + \text{interval}|\text{username}) \tag{4}$$

The data has been checked for normality and homogeneity by visual inspections of a plot of the residuals against the fitted values. Additionally, there was no evidence for non-normality of the data (Shapiro-Wilk test; W=0.97, p>0.05). However, the likelihood ratio test only tells that there is statistical significance for the full model (3). It does not report meaningful results for its individual predictors and interactions. As reported in section 3.5, significance for parameter estimation is possible by providing lower- and upper-bound *p-values*.

Table 2 provides the parameter estimation for the fixed effects (expressed in *z-scores*), the significance of the parameter estimation, and the percentage of the deviance explained by each fixed effect. A single star (*) highlights the significant results (*p-value* less than 0.01). The values have been computed by the *pamer.fnc* function provided by *LMERConvenienceFunctions* [76]. At the 0.01 significance level, valence and dominance are positively correlated with the self-assessed productivity of software developers. Therefore, there is significant evidence to support the hypotheses H1 and H3. Although there is no evidence to support H2, the hypothesis regarding arousal, a possible explanation is proposed in the next section. The random effects are reported in Table 3. The scalar random effects values for the participants belonged to the interval [-0.20, 0.15]; the random effects for the time were all estimated to be zero.

The linear mixed-effects model provided an explanation power of 38.17% in terms of percentage of the deviance explained. Valence was estimated to be 0.04 and dominance 0.55, in terms of *z-scores*. The percentage of the deviance explained by the two effects was 13.81 for valence and 22.42 for dominance, which is roughly the estimation power of the whole model.

## 5. DISCUSSION

### *5.1 Evaluation of results and implications*

The empirical results obtained in this study supported the hypothesized positive correlation between the affective state dimensions of valence and dominance with the self-assessed productivity of software developers. In other words, high happiness with the task and the sensation of having adequate skills are positively correlated with self-assessed productivity.

No support was found for the correlation between arousal and self-assessed productivity. It is suspected that the participants might have misunderstood the correlation's role in the questionnaire. All participants raised questions about the arousal dimension during the questionnaire explanations. A possible explanation of no significant interactions between the affects dimensions and time is that the participants worked on different, independent projects. In addition, the random effects related to time were estimated to be zero, thus non-existing. The full model is still worth reporting with time as fixed and random effect because future experiments with a group of developers working on the same project will likely have significant interactions with time.

The results of this study are in line with the related work reported in section 1.2. Nevertheless, it must be noted that the results, methods, and context of this study are novel and can be compared with those of other studies only theoretically. For example, the results of this study are in line with those of Fisher and Noble [19] where real-time, positive affects (expressed with different constructs than those of this study ) of different types of workers are found to be positively correlated with their productivity. However, the results of this study are not completely in line with those of Khan et al. [14] which on the debugging written tests: (1) induced high valence condition alone did not coincide with high debugging performance, (2) induced high arousal condition alone did coincide with high debugging performance, and (3) induced high arousal and valence conditions together were associated with high debugging performance. Lastly, to the authors' knowledge, our study is the first research studying the correlates between software developers' performance and their self-assessed productivity in their natural, real-software working environments.

The post-task interviews acted as a triangulation to the quantitative findings. Events that visibly reduced the productivity of the participants were captured by the experiment data and were validated by the participants' post-task interviews. Although they were trained in employing the measurement instrument, the participants did not realize that the study was about the assessment of their affective states. On the other hand, the perceived uncertainty of the participants regarding arousal might threaten the validity of the study. However, there is no evidence that the misunderstanding happened, and the suspicion has been reported in this article for more transparency. Additionally, when they were asked about the factors that influenced their productivity, no participants mentioned affects or any synonym related to affects. This further enhances the reliability of the study.

Although not strictly related to this study, the uncertainty of the participants on how they defined their own productivity measurement while developing software suggests that alternative venues in measuring and defining development performance can be pursued. One alternative proposal is the newly discovered concept of "relative perceived productivity measure" built as incremental relative steps of previous estimations.

The *theoretical implication* of this study is that the real-time affects related to a software development task are positively correlated with a programmer's self-assessed productivity.

### 5.2 Threats to validity

This section discusses the limitations of this study. Conclusion, internal, construct, and external validity threats have been mitigated while following the classification provided by Wohlin et al. [27].

*Conclusion validity* threats occur when the experimenters draw inaccurate inference from the data because of inadequate statistical tests. The employed linear mixed-effects models are robust to violations suffered by *ANOVA* methods caused by the multiple dependencies of data (see section 2.1). One threat lies in the limited number of participants (8) who worked for about 90 minutes each. However, the background and skills in the sample were balanced. Due to the peculiarity of the repeated measurements and the analysis method, all 72 measurements are valuable. The linear mixed-effects model is capable of addressing the variability due to individual differences and time effects: the obtained statistical results possessed degrees of freedom between 48 and 64, and F-values above the value of 20 when significance has been reached. It has been shown that repeated measures designs do not require more than seven measurements per individual [79]. Two more measurements have been added in order to be able to remove possible invalid data. Lastly, three hypotheses were tested on the same dataset. While *p-value* adjustments techniques seem, to the authors' knowledge, not suitable for linear mixed-effects models, it can still be reported that the adjusted .05 *p-value* for this study would be 0.05 / 3 = 0.016, while the adjusted .01 *p-value* would be 0.01 / 3 = 0.003. As indicated in Table 2, the *p-values* obtained in this study are less than 0.001 (they are actually even less than 0.0001). Therefore, the results of this study have been obtained with an adjusted $p<0.01$ (Bonferroni correction).

*Internal validity* threats are experimental issues that threaten the researcher's ability to draw inference from the data. Although the experiment was performed in natural settings, the fact the individuals were observed and the lack of knowledge about the experiment contents mitigated social threats to internal validity. A series of pilot studies with the measurement instrument showed that the minimum period to interrupt the participants was about 10 minutes if the case study was focused on a single task instead of longer periods of observations. The designed mitigation measures against internal validity threats were further confirmed through the post-interview data, which showed that the experiment design did not disturb nor negatively influence the productivity and the performance of software developers. However, more data is needed on how to extend the experimental session and the measurement intervals.

*Construct validity* refers to issues with the relation between theory and observation. A construct validity threat might come from the use of the self-assessed productivity. Given the difficulty in using traditional software metrics (the project, the task, and the programming language were random in this study), and that measuring software productivity is still an open problem, self-assessed performance is commonly employed in psychology studies [18,19,23]. Additionally, self-assessed performance is consistent (yet not preferred) to objective measurements of performance [66,67]. There is also the evidence that bias is not introduced by mood in self-reports of performance when individuals alone are being observed [23,66]. The researchers carefully observed the participants during the programming task and this further reduced the risk of bias. Post-task interviews included questions on their choices for productivity levels, which resulted in remarkably honest and reliable answers, as expected. The discussions at the post-interviews with the participants on their self-assessed productivity values reinforced the validity of the data.

*External validity* threats are issues related to improper inferences from the sample data to other persons, settings, and situations. The four professional developers and the four students were sampled using a convenient sampling method. Although this method limits the generalizability of the study, the research is applicable to any software development role with any programming language. Future studies should focus on restricted programming languages, project types, and programmers' experiences. Despite that half of the participants were students, it has been argued that students are the next generation of software professionals; they are remarkably close to the interested population if not even more updated on new technologies [80,81]. Secondly, it can be questioned why the present authors studied software developers working alone on their project. People working in a group interact and trigger a complex, powerful network of affects [82]. Thus, to better control the measurements, individuals working alone have been chosen. However, no participant was forced to limit social connections while working, and the experiment took place in natural settings.

### *5.3 Lessons learned*

The authors of this study learned important lessons from this research experience which are shared in this section. First, the experiment design presented in this study is not suitable for continuous application in the industry. It would be counterproductive to ask software developers to self-assess their affects and their productivity each 10 minutes during their entire working days. The measurement interval employed in this study was the result of a pilot test where the participants agreed that an interval of 10 minutes is suitable for a single session of the experiment only. Future studies should aim to find suitable measurement intervals for longer sessions, e.g., the duration of an iteration of the software development lifecycle.

It is important to explain clearly how the experimental task works to the participants. More importantly, some time should be spent in explaining to them how the measurement instrument works. The freely accessible online appendix [60] reports clear instructions, which were derived from the literature in psychology research. The appendix reports that a paper-version of the survey should be available for enabling discussions and questions about the survey itself. This study showed that the participants do not understand the experiment aims, nor do they get that affects are measured. The authors of this article encourage re-use of the online appendix for organizing similar research. Finally, the authors encourage to perform pre- and post-task interviews with the participants, and to be physically present during the programming task and to keep a research diary.

Some lessons learned about linear mixed-effects models are shared as well. These models are recent, and there is still remarkable discussion on how to employ them. An estimation of statistical significance for mixed effects models is possible via different methods, which will hold different numerical values but the same magnitude. At least in *R*, there is an older implementation of linear mixed-effects models called *nlme.lme*, which provides statistical significance out-of-the-box. However, *nlme.lme* does not handle unbalanced designs. The package has been superseded by *lme4.lmer*, which, on the other hand, does not provide *p-values* out-of-the-box. Likelihood ratio test is a good means to obtain significance testing for *lme4.lmer* model components. However, it should be employed for simple models with few parameters plus their interactions. The method provided by *LMERConvenienceFunctions* [76] is straightforward and should be preferable for more complex models. On the other hand, likelihood ratio tests are still useful for comparing two models in terms of fitting, for example by adding a single effect. When comparing two different models in terms of likelihood ratio test (*R anova* function), it is important to vary one effect only; special attention should be given to random effects, as they should be kept consistent in the whole analysis.

Additionally, when comparing different models with likelihood ratio tests (*anova* function in *R*), it is important to set the *REML* parameter to *F* (false), to indicate that the maximum likelihood method (*ML*) is preferred to the restricted maximum likelihood method (*REML*). Despite the fact that *REML* estimates of standard deviations for the random effects are less biased than corresponding *ML* estimations [83], models with *REML* estimation do not work when comparing models using the likelihood ratio test [75].

Lastly, this article reported the full model (3) of which only two fixed effects were found to be significant (valence and dominance). The full model was kept as the final one because it shows a good example of fitting linear mixed-effects models. The model in (5) could have been the final output for this article, as it only reports the significant fixed effects and the random effects.

$$\text{productivity} \sim \text{valence} + \text{arousal} + (1|\text{participant}) + (0 + \text{time}|\text{participant}) \quad (5)$$

However, while choosing to report a reduced, significant model, it might be useful to check how much it differs from the full model initially chosen. In the case of this study, there was no significant difference between the reduced significant model and the initial full model (*anova* in *R*; $\chi^2(5)=3.09$, $p>0.05$). There was no advantage in reporting a reduced model.

## 6. CONCLUSIONS AND FUTURE WORK

### *6.1 Summary*

For more than thirty years, it has been claimed that software developers are essential when considering how to improve the productivity of the development process and the quality of delivered products. However, little research has been done on how the human aspects of developers have an impact on software development activities. This study echoes a call on employing psychological measurements (erroneously called "psychometrics" in Computer Science related literature) in software engineering research by studying how the affects of software developers - emotions, moods, and feelings – are linked to their programming performance.

This article reports a repeated measures research on the correlation of the affects of software developers and their self-assessed productivity. Eight developers working on their individual projects have been observed. Their affects and their self-assessed productivity were measured on intervals of ten minutes. A linear mixed-effects model was proposed in order to estimate the value of the correlation of the affects of valence, arousal, and dominance, as well as the productivity of developers. The model was able to express about the 38% of the deviance of the self-assessed productivity. Valence and dominance, or the attractiveness perceived towards the development task and the perception of possessing adequate skills, were able to provide nearly all the explanation power of the model.

### *6.2 Impact*

This manuscript provides basic theoretical building blocks on researching the human side of software construction in empirical software engineering: among Khan et al. and Shaw, this is one of the first studies examining the role of the affects of software developers. However, to the authors' knowledge, this is the first study examining the correlation of the affects and the performance of software developers working in natural settings on real-world software projects.

This work performs empirical validation of psychological tests and related measurement instruments in software engineering research. This article shows how to conduct and analyze data with multiple dependencies in the context of repeated within-participant measurements. It proposes the employment of linear mixed-effects models to analyze the data, which have been proven effective in repeated measures designs instead of *rANOVA*. It is also stressed out that *rANOVA* should be avoided in such cases.

Last but not least, this study highlights a substantial body of knowledge in psychology and management research on the affects and their impact on cognitive processing abilities. It presents the most important theories behind affects, their classification, and their measurements, and on the best practices to perform psychological measurements in the context of empirical software engineering.

### *6.3 Future work*

Experiments with more participants performing the same programming task will allow the use of traditional software productivity metrics and improve the generalizability of this study. Moreover, future research can also take this study as a basis and further quantify how much more productive happy developers are in comparison to their less cheerful peers. Mood induction techniques should be employed to study causality effects of affects on the productivity of software developers in order to increase the practical implications of the studies on this topic. The authors of this article intend to conduct qualitative studies as well. In particular, there is a soon-to-be-performed study on what actually makes developers happy, and how affects indicate changes in the developer's performance. Finally, process-based studies on software teams with affects measurements are required in order to understand the dynamics of affects and the creative performance of software teams and organizations. The recent initiative called SEMAT Essence [12] attempts to "refound software engineering based on a solid theory, proven principles, and best practices."[13:46] with a kernel for describing, addressing, and measuring the common *things* of software engineering. These things, named alphas, are opportunity, stakeholders, requirements, software system, team, work, and way-of-working. Despite the fact that some human-related entities are addressed, Essence lacks to incorporate human aspects in the kernel. The present work offers a simple and efficient way to assess the health of the software team members, and it constitutes an excellent candidate for the human part of Essence.

In brief, by studying how software developers perceive their development tasks and how affects correlate to their performance, this study opens up a different, new perspective and approach to investigate the human factors of software development.

## ACKNOWLEDGEMENTS

The authors would like to thank the participants of the experiment and Elena Borgogno, for her tireless help in improving this manuscript. The authors are grateful for the insightful comments offered by five anonymous reviewers, who helped us to improve the article when it was first submitted to *PROFES 2013* and then to the *Journal of Software: Evolution and Process* in the current extended form. Lastly, the authors are thankful for the feedback received by all PROFES 2013 participants.

# TABLES

Table 1. Participants and Projects Details

| ID | Gender | Age | Role | Project | Task | Program. Language | Program. Language Experience | Domain experience |
|---|---|---|---|---|---|---|---|---|
| P1 | M | 25 | PRO | Data collection for hydrological defense | Module for data displaying | Java | HIGH | HIGH |
| P2 | M | 26 | PRO | Research Data Collection & Analysis | Script to analyze data | Python | LOW | HIGH |
| P3 | M | 28 | PRO | Human Resources Manager for a School | Retrieval and display of DB data | Java | HIGH | HIGH |
| P4 | M | 28 | PRO | Metrics Analyzer | Retrieval and sending of metrics | C++ | HIGH | HIGH |
| P5 | F | 23 | STU | Music Editor | Conversion of music score to pictures | C++ | LOW | LOW |
| P6 | M | 20 | STU | Code Editor | Analysis of Cyclomatic Complexity | C++ | LOW | LOW |
| P7 | M | 20 | STU | CAD editor | Single-lined labels on objects | C++ | LOW | LOW |
| P8 | M | 20 | STU | SVG Image Editor | Multiple objects on a circle or ellipse | C++ | HIGH | HIGH |

Table 2. Fixed Effects Estimation

| Fixed Effect | Value | Sum of Square | F-value | Upper *p-value* (64 d.f.) | Lower *p-value* (48 d.f.) | Deviance Explained (%) |
|---|---|---|---|---|---|---|
| valence | 0.04* | 8.65 | 22.09 | 0.000 | 0.000 | 13.81 |
| arousal | 0.07 | 0.19 | 0.49 | 0.487 | 0.489 | 0.30 |
| dominance | 0.55* | 14.04 | 35.86 | 0.000 | 0.000 | 22.42 |
| time | -0.03 | 0.09 | 0.24 | 0.626 | 0.626 | 0.15 |
| valence:time | 0.04 | 0.49 | 1.26 | 0.266 | 0.267 | 0.79 |
| arousal:time | -0.03 | 0.40 | 1.03 | 0.313 | 0.315 | 0.65 |
| dominance:time | 0.01 | 0.03 | 0.08 | 0.785 | 0.785 | 0.05 |

Table 3. Random Effects Estimation

| Random Effect | P1 | P2 | P3 | P4 | P5 | P6 | P7 | P8 |
|---|---|---|---|---|---|---|---|---|
| (1 \| participant) | 0.01 | 0.15 | 0.04 | -0.02 | -0.20 | 0.00 | 0.00 | 0.02 |
| (0 + time \| participant) | 0.00 | 0.00 | 0.00 | 0.00 | 0.00 | 0.00 | 0.00 | 0.00 |

FIGURES

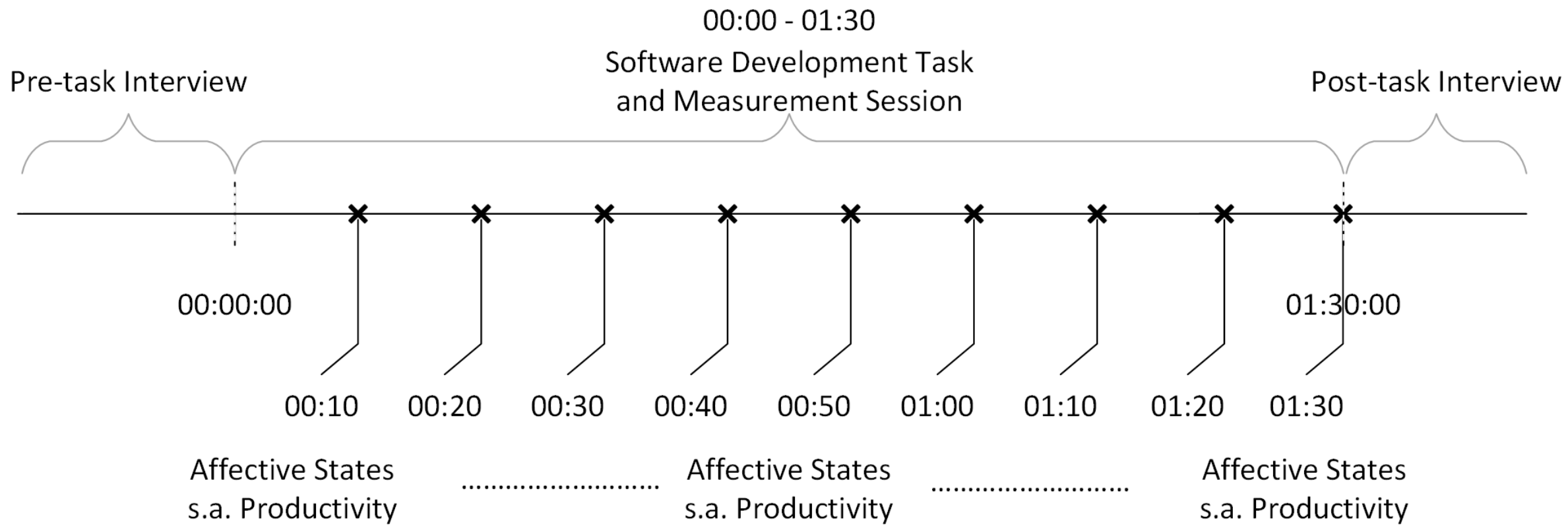

Figure 1. Graphical representation (a timeline) of the research design.

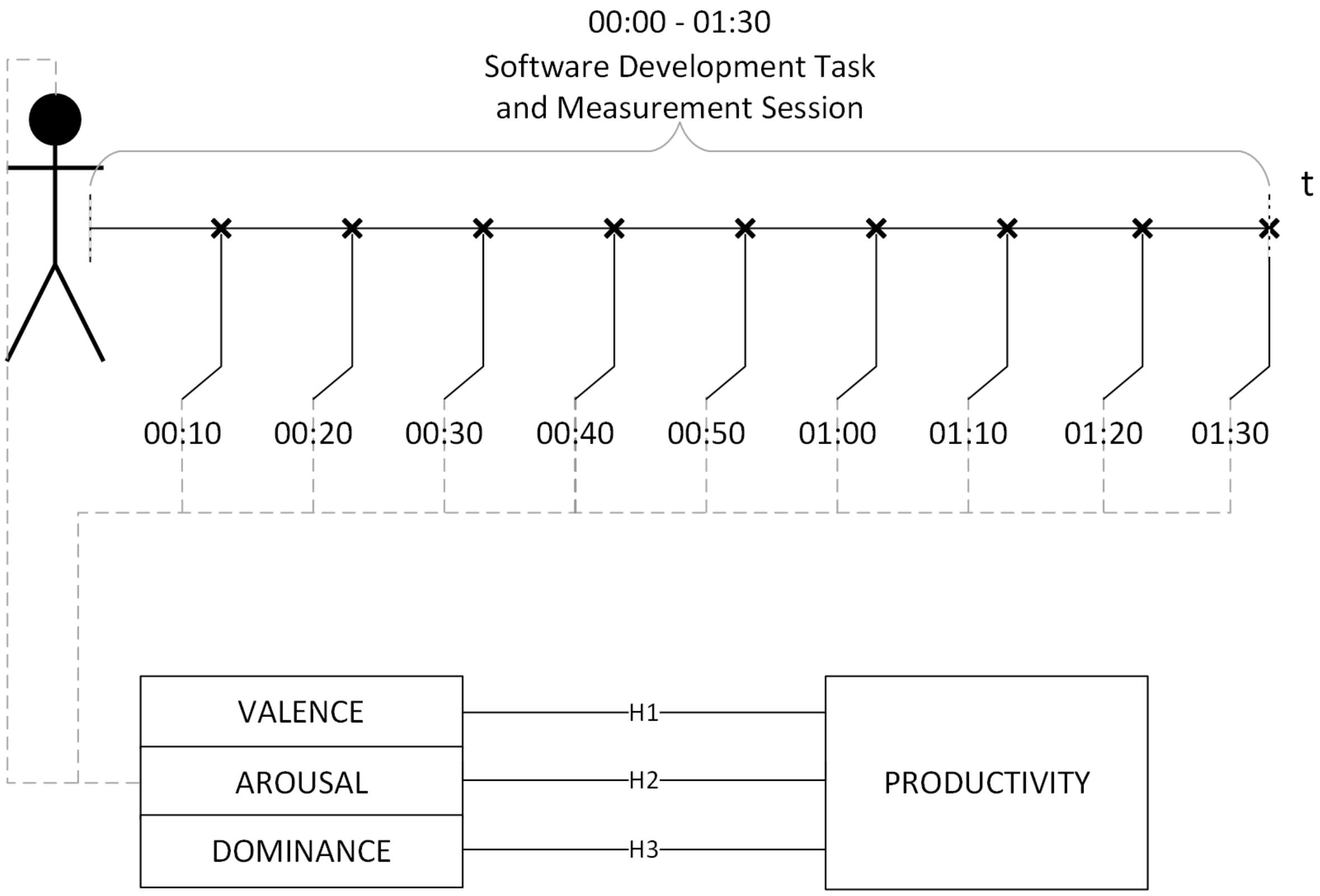

Figure 2. Graphical representation of the hypotheses of this study, the fixed effects, and the random effects in the context of the research design.

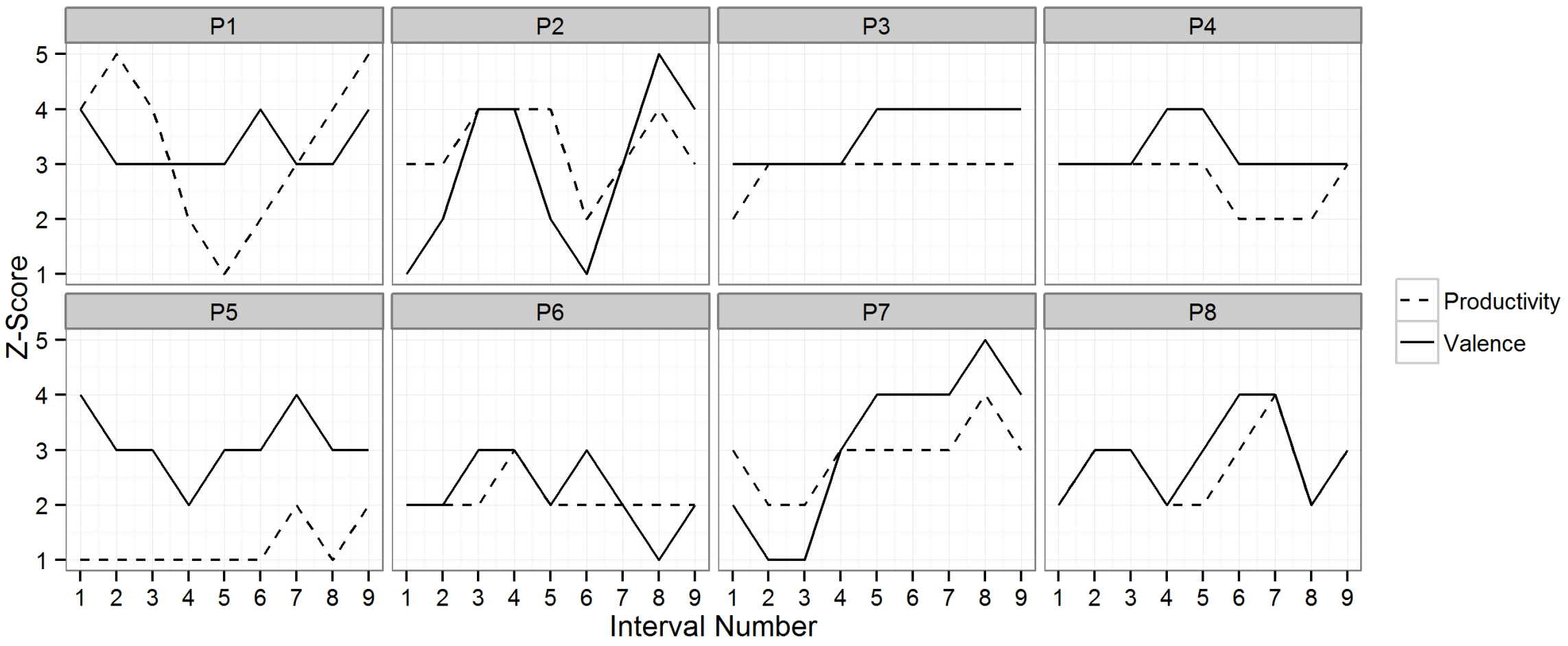

Figure 3. Valence versus productivity over time, grouped per participant.

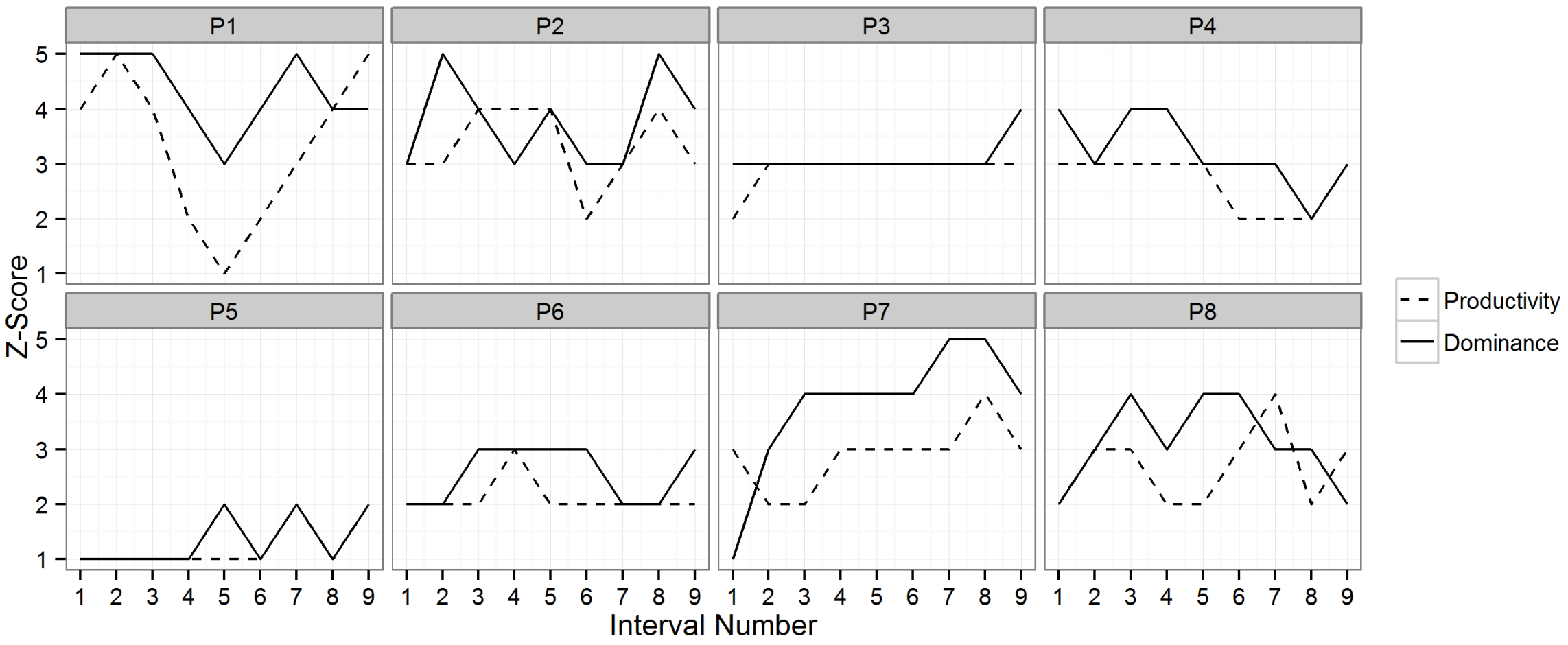

Figure 4. Arousal versus productivity over time, grouped per participant.

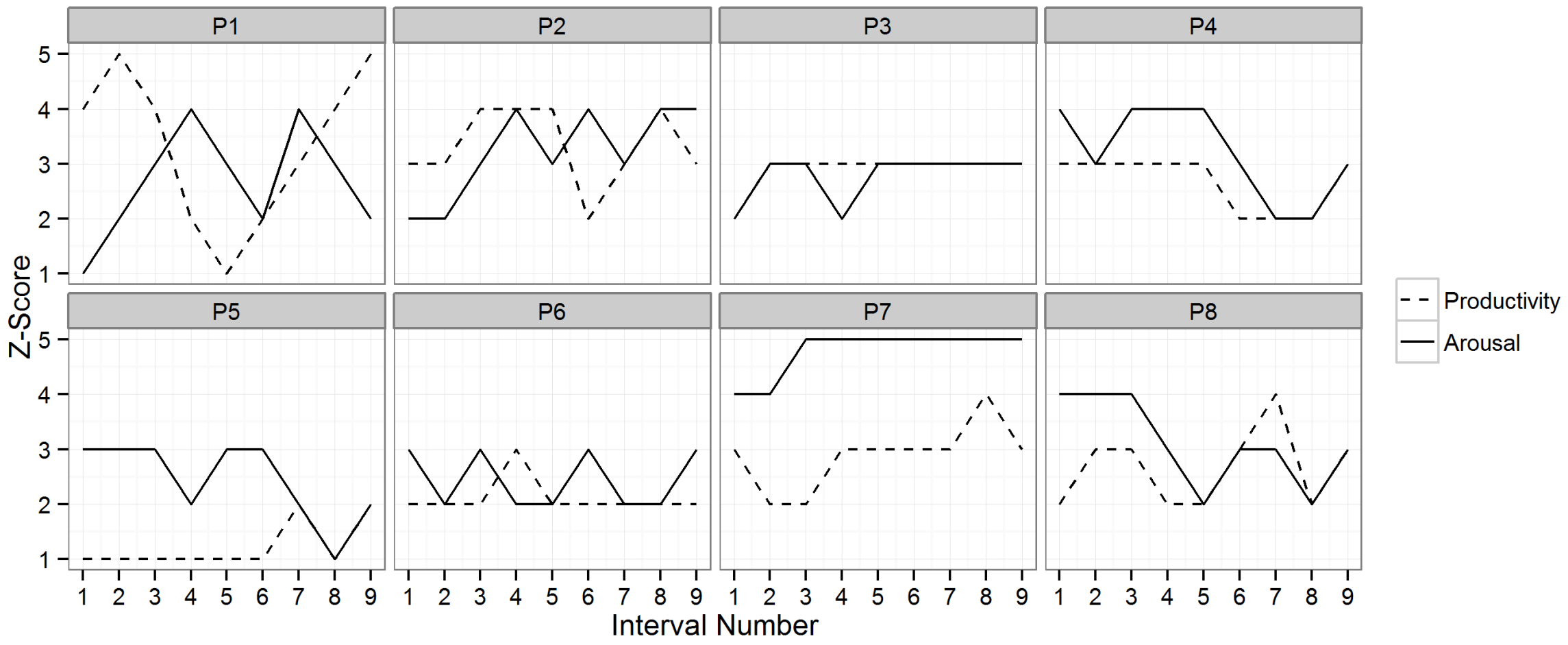

Figure 5. Dominance versus productivity over time, grouped per participant.